\documentclass[%
reprint,
superscriptaddress,
 amsmath,amssymb,
 aps,
pre,
]{revtex4-2}

\usepackage{graphicx}
\usepackage{dcolumn}
\usepackage{bm}

\usepackage{color}
\usepackage{physics}

\newcommand{\NE}{n_{\rm e}}
\newcommand{\NI}{n_{\rm i}}
\newcommand{\NIH}{n_{\rm i1}}
\newcommand{\NIL}{n_{\rm i2}}

\newcommand{\TE}{T_{\rm e}}
\newcommand{\TI}{T_{\rm i}}

\newcommand{\fhot}{f_{\rm i1}}
\newcommand{\fcold}{f_{\rm i2}}

\newcommand{\ve}{v_{\rm e}}
\newcommand{\vi}{v_{\rm i}}

\newcommand{\rci}{r_{\rm ci}}
\newcommand{\rce}{r_{\rm ce}}
\newcommand{\KB}{k_{\rm B}}

\newcommand{\VIN}{v_{\rm in}}
\newcommand{\VOUT}{v_{\rm out}}
\newcommand{\RHOIN}{\rho_{\rm in}}
\newcommand{\RHOOUT}{\rho_{\rm out}}

\newcommand{\SIN}{S_{\rm in}}
\newcommand{\SOUT}{S_{\rm out}}
\newcommand{\KIN}{K_{\rm in}}
\newcommand{\KOUT}{K_{\rm out}}
\newcommand{\HIN}{H_{\rm in}}
\newcommand{\HOUT}{H_{\rm out}}

\newcommand{\TOUT}{T_{\rm out}}
\newcommand{\BIN}{B_{\rm in}}

\newcommand{\NOUT}{n_{\rm out}}
\newcommand{\VA}{v_{\rm A}}

\begin{document}

\preprint{APS/123-QED}

\title{Characterizing the temporal evolution of Biermann-battery-driven magnetic reconnection in laser-ablated plasmas}

\author{T. Morita}
\email{morita@aees.kyushu-u.ac.jp}
\affiliation{Faculty of Engineering Sciences, Kyushu University, Japan}

\author{Y. Muramoto}
\affiliation{Interdisciplinary Graduate School of Engineering Sciences, Kyushu University, Japan}

\author{S. Isayama}
\affiliation{Faculty of Engineering Sciences, Kyushu University, Japan}
\affiliation{Quantum and Spacetime Research Institute (QuaSR), Kyushu University, Japan}
\affiliation{International Research Center for Space and Planetary Environmental Science (i-SPES), Kyushu University, Japan}

\author{M. Edamoto}
\affiliation{Faculty of Science and Technology, Seikei University, Japan}

\author{M. Hanano}
\affiliation{Graduate School of Science, The University of Osaka, Japan}

\author{R. Ishikawa}
\affiliation{Department of Physics and Mathematics, Aoyama Gakuin University, Japan}

\author{Y. Kanesada}
\affiliation{Interdisciplinary Graduate School of Engineering Sciences, Kyushu University, Japan}

\author{K. Koba}
\affiliation{Interdisciplinary Graduate School of Engineering Sciences, Kyushu University, Japan}

\author{H. Kondo}
\affiliation{Department of Physics and Mathematics, Aoyama Gakuin University, Japan}

\author{S. Kurimaru}
\affiliation{Interdisciplinary Graduate School of Engineering Sciences, Kyushu University, Japan}

\author{K. Maeda}
\affiliation{Graduate School of Science, The University of Osaka, Japan}

\author{Y. Maenosono}
\affiliation{Interdisciplinary Graduate School of Engineering Sciences, Kyushu University, Japan}

\author{S. Matsukiyo}
\affiliation{Faculty of Engineering Sciences, Kyushu University, Japan}
\affiliation{Quantum and Spacetime Research Institute (QuaSR), Kyushu University, Japan}
\affiliation{International Research Center for Space and Planetary Environmental Science (i-SPES), Kyushu University, Japan}
\affiliation{Institute of Laser Engineering, The University of Osaka, Japan}

\author{A. Morita}
\affiliation{Department of Physics and Mathematics, Aoyama Gakuin University, Japan}

\author{Y. Nagamatsu}
\affiliation{Interdisciplinary Graduate School of Engineering Sciences, Kyushu University, Japan}

\author{G. Nakayama}
\affiliation{Interdisciplinary Graduate School of Engineering Sciences, Kyushu University, Japan}

\author{T. Ogawa}
\affiliation{Interdisciplinary Graduate School of Engineering Sciences, Kyushu University, Japan}

\author{K. Oshida}
\affiliation{Interdisciplinary Graduate School of Engineering Sciences, Kyushu University, Japan}

\author{Y. Pan}
\affiliation{Faculty of Engineering, Hokkaido University, Japan}

\author{K. Sakai}
\affiliation{National Institute for Fusion Science, Japan}

\author{T. Sano}
\affiliation{Institute of Laser Engineering, The University of Osaka, Japan}

\author{Y. Sato}
\affiliation{Interdisciplinary Graduate School of Engineering Sciences, Kyushu University, Japan}

\author{N. Shimoda}
\affiliation{Department of Physics and Mathematics, Aoyama Gakuin University, Japan}

\author{J. Shiota}
\affiliation{Department of Physics and Mathematics, Aoyama Gakuin University, Japan}

\author{Y. Sudo}
\affiliation{Department of Physics and Mathematics, Aoyama Gakuin University, Japan}

\author{Y. Suzuki}
\affiliation{Graduate School of Science, The University of Osaka, Japan}


\author{T. Takezaki}
\affiliation{Faculty of Engineering, University of Toyama, Japan}

\author{S. J. Tanaka}
\affiliation{Department of Physics and Mathematics, Aoyama Gakuin University, Japan}

\author{K. Tomita}
\affiliation{Faculty of Engineering, Hokkaido University, Japan}

\author{S. Ueno}
\affiliation{Department of Physics and Mathematics, Aoyama Gakuin University, Japan}

\author{S. Yakura}
\affiliation{Department of Physics and Mathematics, Aoyama Gakuin University, Japan}

\author{R. Yamazaki}
\affiliation{Department of Physics and Mathematics, Aoyama Gakuin University, Japan}
\affiliation{Institute of Laser Engineering, The University of Osaka, Japan}

\author{Y. Sakawa}
\affiliation{Institute of Laser Engineering, The University of Osaka, Japan}


\date{\today}

\begin{abstract}
	This paper presents an experimental investigation of magnetic reconnection
	between two laser-produced expanding plasmas,
	focusing on the quantitative evaluation of the reconnection rate
	and energy conversion under varying initial conditions.
	By changing the separation distance between the drive laser focal spots
	(1 mm and 2 mm), we systematically controlled the inflow parameters.
	The reconnection region was probed using a two-directional
	laser Thomson scattering (LTS) system,
	which simultaneously measured the local plasma parameters
	parallel to the outflows and along the current sheet.
	Based on our established method incorporating
	macroscopic energy and mass conservation laws
	to derive the upstream magnetic field directly from LTS spectra,
	we characterized the temporal evolution of the current sheet
	and the reconnection rate.
	The larger spot separation allows the plasma bubbles to
	expand for a substantially longer time before the formation
	of a reconnection current sheet, resulting in different
	upstream conditions. Nevertheless, both configurations
	yielded comparable upstream magnetic fields and
	reconnection rates. These results suggest that the
	reconnection rate is relatively insensitive to the global
	inflow conditions and is primarily controlled by the local
	physics of the reconnection layer once a current sheet is
	formed.
\end{abstract}

\maketitle

\section{\label{intro}Introduction}

Magnetic reconnection is a fundamental plasma process that involves
the rapid reconfiguration of magnetic field topology,
leading to the rapid conversion of stored magnetic energy
into plasma kinetic and thermal energies.
This phenomenon plays a significant role in various astrophysical environments,
such as solar flares, Earth's magnetospheric substorms,
and the dynamics of 
accretion disks\cite{Yamada2010-jd,Zweibel2009-ci,Burch2016-et,Hesse2020-dl}.
In recent years, high-power laser systems have provided a controllable
laboratory platform to investigate the micro- and macro-physics of
magnetic reconnection under high-energy-density (HED)
conditions.
Various experimental approaches have been developed to investigate reconnection dynamics.
These include magnetic reconnection between colliding plasmas in external magnetic fields\cite{Fiksel2014-mp},
the investigation of pure electron reconnection in a weak external magnetic field
where only electrons are magnetized\cite{Kuramitsu2018-ja,Sakai2022-uu},
strongly driven reconnection using laser-driven capacitor coils\cite{Zhang2023-tf,Chien2023-yp},
and relativistic reconnection experiments utilizing kilo-tesla scale magnetic fields\cite{Law2020-wk}.
In contrast, experiments utilizing the collision of two expanding plasma bubbles,
each carrying a self-generated azimuthal magnetic field
induced by the Biermann battery effect\cite{Stamper1971-uf,Haines1997-ef,Campbell2022-dn},
offer the advantage of a simple experimental setup,
where reconnection can be driven merely by irradiating two focal spots with lasers\cite{Nilson2006-aj,Li2007-hy,Zhong2010-ko,Rosenberg2015-cj,Tubman2021-cb,Fiksel2021-zs,Zhao2022-ax,Ping2023-qx,Valenzuela-Villaseca2024-pc}.
However, detailed measurements of the inflow and outflow dynamics in such configurations
are difficult because the phenomena occur within a microscopic spatial scale of a few hundred micrometers
and a short time duration of a few nanoseconds.

While previous laboratory studies have demonstrated
the formation of current sheets and the subsequent generation of
plasma outflows, capturing the temporal evolution
of the reconnection rate and quantifying the energy partition
between thermal and kinetic energies remain
challenging\cite{Dong2012-kc,Fox2011-rt}.
Specifically, simultaneous and localized measurements of plasma parameters
along both the outflow and the current sheet directions are essential
to evaluate the conversion efficiency of magnetic energy.
In our previous study\cite{Morita2022-cm},
we established a diagnostic technique to derive
the upstream magnetic field and the reconnection rate directly
from laser Thomson scattering (LTS) spectra,
eliminating the need for external magnetic probes.

Building upon this established framework, this paper presents
a comparative experimental study of 
the dynamics and temporal evolution of magnetic reconnection
under different initial inflow conditions.
By varying the separation distance between the two drive laser focal spots
(1 mm and 2 mm),
we systematically modified the interaction parameters of the colliding plasmas.
Using a two-directional LTS system\cite{Sheffield2010-hg,Morita2022-cm},
we simultaneously measured the plasma properties to track
the temporal evolution of the magnetic fields, current sheet dissipation,
and outflow generation.
The primary objective is to quantitatively evaluate
the magnetic reconnection rate, the energy transfer from the magnetic field
to the plasma thermal and kinetic components, and the onset timing
of the reconnection process across the different inflow conditions.

An open question in magnetic reconnection is how
strongly the reconnection dynamics depend on the
global geometry of the system.
In laser-driven Biermann-battery reconnection, the
upstream magnetic flux and inflow conditions are
expected to vary with the initial focal-spot
separation.
It remains unclear, however, whether 
the reconnection rate itself is similarly sensitive to
such changes in the global geometry and inflow conditions.
In this work, we address this question through
temporally resolved and simultaneous measurements
of the current sheet and reconnection outflow.
Our measurements show that the two focal-spot separations
lead to substantially different plasma expansion histories
prior to current-sheet formation and therefore are expected
to produce different upstream conditions.
Nevertheless, comparable reconnection rates are obtained
once a reconnection current sheet is formed.

The structure of this paper is as follows.
Section II describes the experimental setup at the Gekko-XII laser facility,
including the self-emission and two-directional LTS diagnostics.
Section III presents the measurement results,
focusing on the LTS spectra and the macroscopic formation of the current sheet.
Section IV provides a detailed discussion on the derivation of
local plasma parameters, the evaluation of the reconnection rate,
and a theoretical analysis of the energy balance and outflow dynamics.
Finally, Section V summarizes our results and discussion.

\section{Experimental setup}

\begin{figure}
\begin{center}
\includegraphics[width=\linewidth]{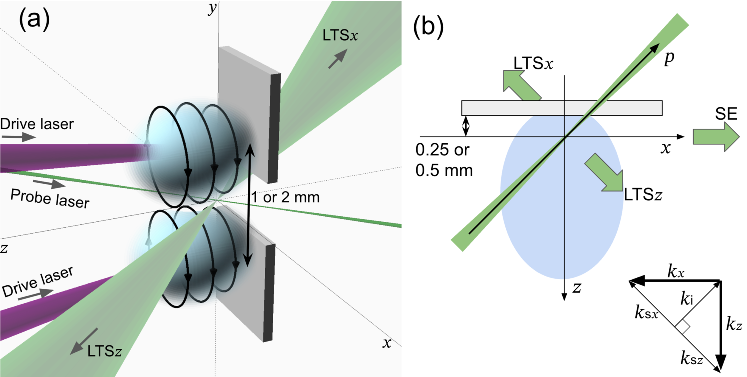}
\caption{\label{fig:setup}
(a) Experimental setup with two planar foils,
	two drive laser beams, and a probe laser for LTS measurements.
	The self-generated magnetic field is illustrated as black arrows.
	(b) The top view of the experimental setup with
	the probe laser and scattered light on the $x$-$z$ plane 
	at 90 degrees.
	Two-direction measurements are shown:
	$\text{LTS}_x$ for outflow measurement and
	$\text{LTS}_z$ for current sheet measurement.
	The self-emission (SE) is imaged and measured with a streak camera
	along the $x$-direction.
}
\end{center}
\end{figure}

The experiment was conducted using the Gekko-XII laser facility
at the Institute of Laser Engineering, the University of Osaka.
Two Gekko drive laser beams irradiated two planar carbon foils
to produce expanding plasma bubbles as shown in Fig. \ref{fig:setup}(a).
The separation distance between the two laser focal spots was set to either
1~mm or 2~mm to systematically vary the inflow parameters and investigate
the dependence of the reconnection process.
The drive lasers delivered an energy of $\sim$600~J per beam
with a pulse duration of $\sim$1.3~ns
(Gaussian in space and time) at a wavelength of 1053~nm.
The focal spot diameter was approximately 100~$\mu$m,
resulting in an on-target intensity of $\sim 5.9 \times 10^{15}$~W/cm$^2$.
The magnetic field is generated through 
the Biermann battery mechanism, which arises when the electron density 
and temperature gradients are not parallel\cite{Stamper1971-uf,Haines1997-ef,Campbell2022-dn}. 
In this case,
$\partial \vb*{B}/\partial t=(\KB/e\,\NE)
(\vb*{\nabla}\TE\times\vb*{\nabla}\NE)$,
where $\KB$ is the Boltzmann constant, $e$ is the
elementary charge, and $\NE$ is the electron density.
This relation indicates that the field-generation rate
increases with the density and temperature gradients.
In laser-produced plasmas, the density gradient is mainly directed
normal to the target surface, whereas the temperature gradient has a
radial component away from the focal spot.
Their cross product therefore generates an azimuthal magnetic field
surrounding the plasma bubble.
The two magnetized plasmas interacted with each other at the $y\sim0$ plane,
reconnecting the anti-parallel magnetic fields to heat and accelerate
the inflow plasmas and to form outflows in the $\pm x$ directions.

The self-emission (SE) from the two plasma bubbles
expanding from the top and bottom was imaged
with a streak camera along the $x$-direction.
The entrance slit of the streak camera was arranged along the $y$-axis at $z=0$
to observe the temporal evolution of the interaction between 
the two plasmas.

To directly probe the magnetic reconnection dynamics, we employed
a two-directional laser Thomson scattering (LTS) system.
A separate probe laser with a wavelength of 532~nm, an energy of $\sim$300~mJ,
and a pulse duration of $\sim$5~ns was injected into the midplane
between the two expanding plasmas.
The probe laser propagating along the $p$-axis
and the collected scattered light were configured
on the $x$-$z$ plane at a 90-degree angle as shown in Fig. \ref{fig:setup}(b).
This diagnostic setup enabled the simultaneous measurement of
local plasma parameters in two orthogonal directions.
Specifically, the scattered light was divided into two paths:
the $\text{LTS}_x$ channel, which measured the plasma properties
parallel to the outflows,
and the $\text{LTS}_z$ channel, which measured the properties
along the current sheet.
This geometry is illustrated by the vector relations in Fig. \ref{fig:setup}(b).
$k_{\rm i}$, $k_{{\rm s}x}$, and $k_{{\rm s}z}$ represent the wave vectors for the incident laser and the scattered light in the $x$ and $z$ directions, respectively.

The scattered signals were dispersed by spectrometers and recorded by
intensified CCD (ICCD) cameras with a gate width of 3~ns for both channels,
providing a detailed view of the temporal evolution of the reconnection region.
Measurements were taken at various delay times ranging from 4 to 6~ns
relative to the main drive laser pulse for the 1-mm separation and
from 8 to 12~ns for the 2-mm separation, respectively,
to capture the formation and dissipation of the current sheets 
using $\text{LTS}_z$
and the heated and accelerated outflows using $\text{LTS}_x$.

\section{Results}

\begin{figure}
\begin{center}
\includegraphics[width=\linewidth]{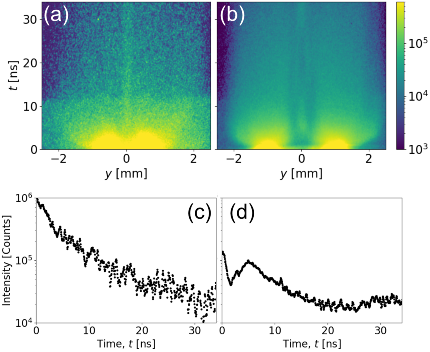}
\caption{\label{fig:sop}
Temporal evolution of self-emission intensity along the $y$-axis at $z=0$
	for (a) 1-mm and (b) 2-mm separations.
	The line-out plots at $y=0$ mm for (c) 1-mm and (d) 2-mm separations, 
	respectively.
}
\end{center}
\end{figure}

Figures \ref{fig:sop}(a) and \ref{fig:sop}(b) show the self-emission
images recorded with a streak camera
for the 1-mm and 2-mm separations, respectively.
The horizontal and vertical axes show the $y$-position and time, respectively.
Generally, the emission is interpreted as bremsstrahlung emission,
which strongly depends on the electron density\cite{Rybicki1980-my}.
Figures \ref{fig:sop}(c) and \ref{fig:sop}(d) are the line-out plots
at $y=0$ mm for 1-mm and 2-mm separations, respectively.
For the 1-mm separation, the two plasmas interacted at an early stage
(likely at $t<4$ ns).
However, it is difficult to resolve the intensity (density) change
due to limited time resolution and signal saturation.
On the other hand, for the 2-mm separation shown in Fig. \ref{fig:sop}(b),
the density change is imaged:
two plasma bubbles interact with each other at $t\sim 5$ ns,
showing a density increase, followed by a density decrease 
from $t\sim5$ to 15 ns,
suggesting plasma exhaust from the midplane.

\begin{figure}
\begin{center}
\includegraphics[width=\linewidth]{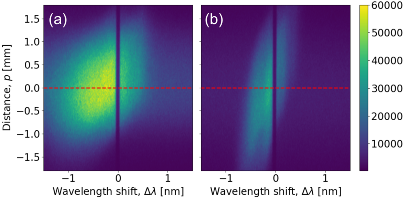}
\caption{\label{fig:TSz2D}
	Thomson scattering spectra in the $z$-direction $\text{LTS}_z$ for
	(a) the 1-mm and (b) 2-mm beam separations
	at $t=4$ and 10 ns, respectively.
	The horizontal and vertical axes show the wavelength relative
	to the incident laser wavelength, $\Delta\lambda$,
	and distance along the probe laser, $p$, respectively.
}
\end{center}
\end{figure}

Figures \ref{fig:TSz2D}(a) and \ref{fig:TSz2D}(b) show
Thomson scattering spectra
in the z-direction ($\text{LTS}_z$)
for the 1-mm and 2-mm beam separations at $t = 4$ and $10$ ns,
respectively.
The horizontal and vertical axes show the wavelength relative to
the incident laser wavelength, $\Delta\lambda$,
and distance along the probe laser, $p$, respectively.
Line-out plots were extracted from these spectra to investigate
the local plasma conditions within the reconnection region.

\begin{figure}
\begin{center}
\includegraphics[width=\linewidth]{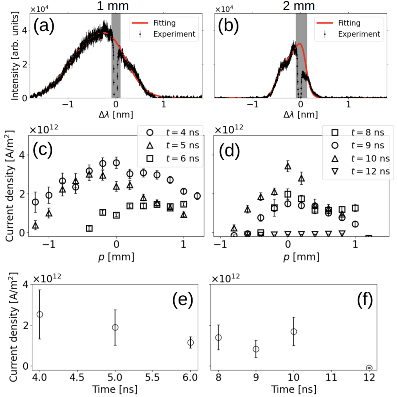}
\caption{\label{fig:TSz}
The line-out plots of LTS spectra (a) at $t=4$ ns and $p=0$ mm,
	and (b) at $t=10$ ns and $p=0$ mm.
	The current density is estimated for both targets [(c) and (d)]
	at different positions along $p$.
	Panels (e) and (f) show the averaged current density estimated
	from multiple laser shots at different times and
	plotted as a function of time.
}
\end{center}
\end{figure}
Figures \ref{fig:TSz}(a) and \ref{fig:TSz}(b) show the line-out plots of
LTS spectra
at the center of the current sheet ($p = 0$ mm) at $t = 4$ ns and $t = 10$ ns
for the 1-mm and 2-mm separations, respectively.
The signal intensity around $\Delta\lambda \sim 0$ (shaded area) 
is strongly damped by a notch filter introduced in the spectrometer.
Both spectra show asymmetric profiles with different heights 
for red- and blue-shifted resonant peaks,
suggesting different Landau damping on the ion-acoustic waves
in the $\pm k$ directions,
resulting from the relative drift velocity between electrons and ions.
By considering the electron drift and electron densities inferred
from the fitting with a spectral density function\cite{Sheffield2010-hg},
the current density is estimated for both targets at different positions
along $p$.

From multiple laser shots with the same targets at $t=4$, 5, 6 ns for
the 1-mm separation, and at $t=8$, 9, 10, 12 ns for the 2-mm separation, respectively,
time- and space-resolved current densities are calculated
as shown in Figs. \ref{fig:TSz}(c) and \ref{fig:TSz}(d),
with the relation,
$j = Ze\NI\vi - e\NE\ve$,
where $Z$ is the averaged charge state, $e$ is the elementary charge,
$\NI$ and $\NE$ are the ion and electron densities, respectively,
and $\vi$ and $\ve$ are the ion and electron flow velocities, respectively.
All these parameters are obtained from the spectral fitting shown
in Figs. \ref{fig:TSz}(a) and \ref{fig:TSz}(b),
and the error bars on the current density are evaluated
from the standard errors on the densities and temperatures.
The spatially averaged current density, estimated from multiple laser shots,
is plotted as a function of time in Figs. \ref{fig:TSz}(e) and \ref{fig:TSz}(f)
to observe the macroscopic formation and dissipation of the current sheet.
Figure \ref{fig:TSz}(e) for the 1-mm separation indicates
a decay in the current density from 4 to 6 ns,
while Fig. \ref{fig:TSz}(f) for the 2-mm separation shows
a large current density from 8 to 10 ns followed by a decay from 10 to 12 ns.

\begin{figure*}
\begin{center}
\includegraphics[width=\linewidth]{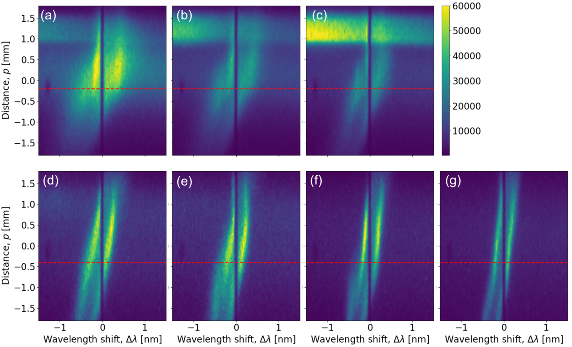}
\caption{\label{fig:TSx2D}
	Thomson scattering spectra in the $x$-direction ($\text{LTS}_x$)
	for 1-mm separation
	at (a) $t=4$, (b) 5, and (c) 6 ns, and
	for 2-mm separation at (d) 8, (e) 9, (f) 10, and (g) 12 ns.
}
\end{center}
\end{figure*}

Figure \ref{fig:TSx2D} presents the $\text{LTS}_x$ spectra
over extended time intervals to capture the temporal evolution of
the entire process in the outflow direction:
at $t = 4, 5,$ and $6$ ns for the 1-mm separation,
and at $t = 8, 9, 10$, and $12$ ns for the 2-mm separation.
As shown in the spectra for the 1-mm separation
[Figs. \ref{fig:TSx2D}(a)--\ref{fig:TSx2D}(c)]
and for the 2-mm separation at $t=8, 9$, and 10 ns
[Figs. \ref{fig:TSx2D}(d)--\ref{fig:TSx2D}(f)],
each spectrum shows double peaks with different peak widths and intensities.
As discussed in our previous study\cite{Morita2022-cm}
and observed in other colliding plasma experiments\cite{Ross2010-qo},
these spectra in the $x$-direction are well fitted with
a spectral density function with two ion components.
Unlike the LTS$_z$ spectra, which are adequately reproduced by a
single-Maxwellian ion distribution, the LTS$_x$ spectra exhibit
differences not only in peak intensity but also in spectral width.
A single drifting Maxwellian can reproduce the intensity asymmetry
arising from the relative drift between electrons and ions, 
but it cannot simultaneously account
for the observed differences in both peak intensity and peak width.
A double-Maxwellian ion distribution is therefore introduced as the
minimum model required to reproduce the observed spectral features.

\begin{figure}
\begin{center}
\includegraphics[width=\linewidth]{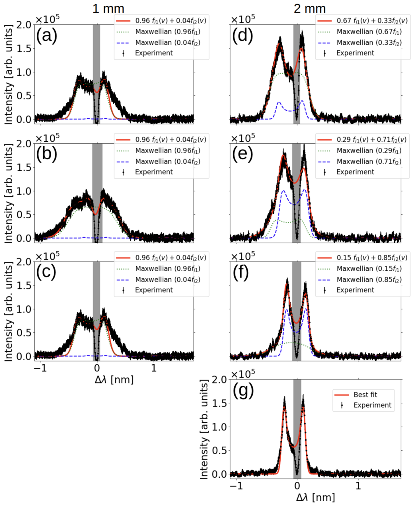}
\caption{\label{fig:TSx}
	The line-out plots of LTS spectra
	(dashed lines in Fig. \ref{fig:TSx2D})
	at $p=-0.2$ mm at
	(a) $t=4$,
	(b) 5, and
	(c) 6 ns for the 1-mm separation,
	and spectra at $p=-0.4$ mm at
	(d) $t=8$ ns,
	(e) 9,
	(f) 10, and
	(g) 12 ns for the 2-mm separation.
}
\end{center}
\end{figure}

Line-out plots of LTS spectra at slightly offset positions
(e.g., $p = -0.2$ mm for the 1-mm separation and 
$-0.4$ mm for the 2-mm separation,
red dashed lines shown in Fig. \ref{fig:TSx2D}) are
presented in Fig. \ref{fig:TSx} for analyzing the interaction regions
with inflow and outflow.
These spectra were fitted with a spectral density function
that includes a single Maxwellian for the electron velocity distribution
and a double Maxwellian for the ion velocity distribution.
The solid lines show the best fit results with
hot and cold ion velocity distributions $\fhot$ and $\fcold$,
respectively, as the total ion velocity distribution
$f(v) = (\NIH/\NI)\fhot + (\NIL/\NI)\fcold$,
where $\NIH$ and $\NIL$ are the densities of the hot and cold components, respectively.
On the other hand, dashed and dotted lines
show the spectra assuming single ion velocity distributions of 
hot ($\fhot(v)$, dotted lines) and 
cold ($\fcold(v)$, dashed lines) components
with the same plasma parameters as the best fit, respectively.
Although more complicated distributions involving three or more ion
populations may also be possible, the present data do not uniquely
constrain such models.
Because magnetic reconnection naturally produces distinct inflow and
outflow populations, we adopt the simplest physically motivated
two-component description.

\section{Discussion}

The present analysis is based on the methodology
developed in Ref.~\cite{Morita2022-cm}.
The LTS$_z$ spectra probe the current-sheet region
and provide the electron drift velocity and current
density $j$ through analysis of the ion-acoustic
peak asymmetry.
The LTS$_x$ spectra are analyzed using hot and cold
ion components, which are interpreted as
reconnection outflow and inflow plasmas,
respectively, and provide the characteristic
outflow velocity $\VOUT$.
These experimentally derived quantities are then
incorporated into energy and mass conservation
relations to estimate the upstream magnetic field
$\BIN$, magnetic flux $\phi$, and reconnection rate $R$.

Using this framework, the present study compares two target
configurations with different focal-spot separations (1 mm and 2 mm)
to investigate how the temporal evolution of reconnection depends on
the initial inflow conditions.

\subsection{Energy Balance and Outflow Dynamics}

The energy conversion from the magnetic field to the plasma
was quantitatively validated by considering 
the energy and mass conservation laws
across the magnetic reconnection layer, based on a standard steady-state
macroscopic reconnection framework \cite{Priest2000, Zweibel2009-ci}.
The macroscopic energy balance between the inflowing electromagnetic energy and the outflowing plasma energy can be expressed as:
\begin{align}
	\SIN L \VIN &= \left( \KOUT + \HOUT \right) \delta \VOUT, \label{eq:energy_balance}
\end{align}
where $L$ and $\delta$ are the length and width of the current sheet,
respectively.
The energy fluxes across the reconnection region are evaluated using the Poynting flux ($S v$), the kinetic energy flux ($K v$), and the enthalpy flux ($H v$).
Here, the effective electromagnetic energy density $S = B^2/\mu_0$ accounts for both the advection of magnetic energy and the work done by magnetic pressure:
\begin{align}
	S v &= (B^2/\mu_0) v, \label{eq:poynting}\\
	K v &= \frac{1}{2}\rho v^3, \label{eq:kinetic}\\
	H v &= (u+p)v = \frac{\gamma}{\gamma-1}pv = \frac{\gamma}{\gamma-1}nTv, \label{eq:enthalpy}
\end{align}
where $B$ is the magnetic field, $\mu_0$ is the vacuum permeability, $\rho$ is the mass density, $\gamma = 5/3$ is the adiabatic index assuming an ideal gas,
$u = p/(\gamma-1)$ is the internal energy density,
$p = nT$ is the plasma pressure, $T$ is the temperature,
and $n$ is the plasma density.
Here, the kinetic and enthalpy fluxes in the inflow are assumed to be zero
($\KIN = \HIN = 0$) due to deceleration by magnetic pressure and
the relatively low temperature and density of the inflow.
Furthermore, the Poynting flux in the outflow is assumed to be zero ($\SOUT=0$),
indicating efficient energy conversion.
Similarly, mass conservation across the reconnection region requires that the mass inflow equals the mass outflow:
\begin{align}
	\RHOIN L \VIN &= \RHOOUT \delta \VOUT.
\label{eq:mass_conservation}
\end{align}

We introduce the parameter $\alpha$, which characterizes the ratio of
the thermal energy density to the kinetic energy density, to relate
the thermal and kinetic energies measured in the outflow:
\begin{align}
	\alpha = \frac{\NOUT\TOUT}{\KOUT}.
\label{eq:alpha}
\end{align}

By substituting the flux definitions
(Eqs. \ref{eq:poynting}--\ref{eq:enthalpy})
and the energy ratio $\alpha$ into the energy balance equation
(\ref{eq:energy_balance}),
we obtain the relationship between the upstream magnetic field
and the outflow conditions:
\begin{align}
	\frac{\BIN^2}{\mu_0} L \VIN &= \left( \frac{1}{2}\RHOOUT\VOUT^2 + \frac{\gamma}{\gamma-1}\NOUT\TOUT \right)\delta\VOUT \nonumber \\
	&= \left( \frac{\alpha\gamma + \gamma - 1}{2(\gamma-1)}\RHOOUT \VOUT^2 \right) \delta \VOUT.
\label{eq:energy_substituted}
\end{align}

Using Eq. (\ref{eq:energy_substituted}) alongside
the mass conservation equation (\ref{eq:mass_conservation}),
the outflow velocity $\VOUT$ can be expressed as a function of
the upstream Alfv\'{e}n velocity $\VA = \BIN/\sqrt{\mu_0 \RHOIN}$:
\begin{align}
	\VOUT &= \sqrt{\frac{2(\gamma-1)}{\alpha\gamma + \gamma - 1}}\frac{\BIN}{\sqrt{\mu_0 \RHOIN}} \label{eq:vout} \\
	&= \sqrt{\frac{2(\gamma-1)}{\alpha\gamma + \gamma - 1}}\VA \nonumber
\end{align}

Conversely, by rearranging Eq. (\ref{eq:vout}),
the upstream magnetic field $\BIN$ can be estimated
directly from the experimentally measured outflow
velocity $\VOUT$ and the energy partition
parameter $\alpha$:
\begin{align}
	\BIN = \sqrt{\frac{\mu_0 \RHOIN (\alpha\gamma + \gamma - 1)}{2(\gamma-1)}} \VOUT.
\label{eq:b0}
\end{align}
This framework confirms that the kinetic and thermal energies 
gained by the outflows originate entirely from the dissipated magnetic energy,
providing a theoretical foundation for interpreting our 
temporary and spatially resolved LTS measurements.

\begin{figure}
\begin{center}
\includegraphics[width=\linewidth]{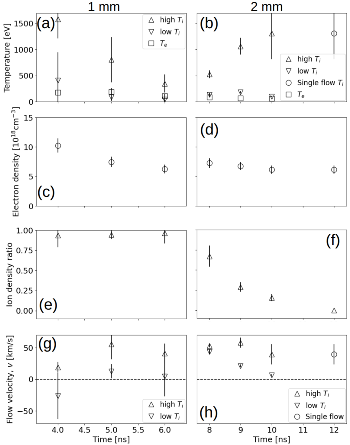}
\caption{\label{fig:params}
	(a) [(b)] Temperatures, 
	(c) [(d)] electron density,
	(e) [(f)] ion density ratio of the hot component to the total density ($\NIH/\NI$),
	and (g) [(h)] flow velocities
	for the 1-mm [2-mm] separation, respectively.
}
\end{center}
\end{figure}

By applying the spectral fitting technique to the line-outs shown
in Fig. \ref{fig:TSx},
we extracted the plasma parameters of the hot and cold components
(Maxwellian velocity distributions of $\fhot$ and $\fcold$, respectively)
for the 1-mm 
[Figs. \ref{fig:params}(a), \ref{fig:params}(c), \ref{fig:params}(e), and \ref{fig:params}(g)] 
and
2-mm separations 
[Figs. \ref{fig:params}(b), \ref{fig:params}(d), \ref{fig:params}(f), and \ref{fig:params}(h)].

It should be noted that, in the present experimental setup,
the inflow and outflow are directed in the $\pm y$ and $\pm x$ directions, 
respectively.
Since LTS measures the $x$ and $z$ directions with
$\text{LTS}_x$ and $\text{LTS}_z$, respectively,
there is no direct measurement of the inflow velocity in the $y$ direction.
Here, we assume that the inflow is strongly decelerated 
($v_{{\rm in},y}\sim0$), and
the effective outflow velocity $\VOUT$ is evaluated from the difference between 
the flow velocity of the hot component and the $x$-component of the cold component from $\text{LTS}_x$.

As shown in Figs. \ref{fig:params}(a) and \ref{fig:params}(b),
the ion temperatures for the cold components are comparable to or slightly larger
than the electron temperature, while the hot components show 
higher ion temperatures.
This suggests that the cold and hot components correspond to the
inflow and outflow, respectively.
While the ion temperature of the hot component decreases over time for the 1-mm separation,
it increases from 8 to 10 ns for the 2-mm separation.
The electron densities [Figs. \ref{fig:params}(c) and \ref{fig:params}(d)]
decrease over time for both the 1-mm and 2-mm separations.

However, the density ratio of the hot component to the total ion density ($\NIH/\NI$, where $\NI = \NIH + \NIL$)
shows different tendencies for the 1-mm and 2-mm separations [Figs. \ref{fig:params}(e) and \ref{fig:params}(f)].
For the 1-mm separation, the ratio remains close to unity from 4 to 6 ns,
suggesting that magnetic reconnection occurs continuously during this period.
In contrast, for the 2-mm separation, the ratio decreases from 8 to 10 ns.
At $t=12$ ns, 
the spectrum [Fig. \ref{fig:TSx}(g)]
exhibits a symmetric profile, which can be well fitted by a single Maxwellian
ion velocity distribution.
Therefore, the plasma parameters at $t=12$ ns are evaluated assuming a single ion component,
as shown in Figs. \ref{fig:params}(b), \ref{fig:params}(d), and \ref{fig:params}(h).
Consequently, a density ratio of zero is plotted at
this time in Fig.~\ref{fig:params}(f), indicating
that the density of the outflow component is also
zero.
The Biermann magnetic field is generated during the
early expansion phase of the laser-ablated plasma,
when strong density and temperature gradients are
present near the laser focal spot.
As the plasma expands away from the focal region,
these gradients rapidly decrease and the
magnetic-field generation rate becomes negligible.
The pre-existing magnetic field is then advected
with the expanding plasma.
As reconnection proceeds, the available magnetic flux
is consumed and expelled from the
interaction region between 8 and 10 ns.
Consequently, the reconnection outflow disappears and
the plasma relaxes toward a stagnated state
[Fig.~\ref{fig:params}(f)], resulting in the
disappearance of the hot component at $t=12$ ns.
These observations indicate that the magnetic
reconnection process has terminated by
$t=12$ ns.
In addition, the hot components exhibit higher flow velocities
than the cold components for both the 1-mm and 2-mm separations,
as shown in Figs. \ref{fig:params}(g) and \ref{fig:params}(h), respectively,
indicating an increase in the bulk kinetic energy.

\begin{figure}
\begin{center}
\includegraphics[width=\linewidth]{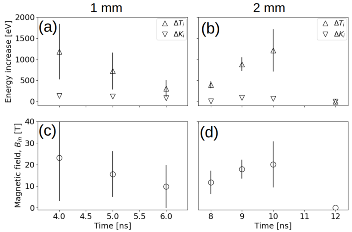}
\caption{\label{fig:b0}
	(a) [(b)] The increments of thermal and kinetic energies
	are calculated 
	from the differences in temperatures and velocities
	between the hot and cold components shown 
	in Fig. \ref{fig:params}(a) [Fig. \ref{fig:params}(b)]
	and Fig. \ref{fig:params}(g) [Fig. \ref{fig:params}(h)],
	for 1-mm [2-mm] separation.
	The upstream magnetic fields [(c) and (d)] are also estimated 
	from Eq. (\ref{eq:b0}).
}
\end{center}
\end{figure}

The increases in thermal and kinetic energies for different times
[Figs. \ref{fig:b0}(a) and \ref{fig:b0}(b)]
are estimated from the differences in ion temperatures 
[Figs. \ref{fig:params}(a) and \ref{fig:params}(b)] 
and flow velocities [Figs. \ref{fig:params}(g) and \ref{fig:params}(h)] 
between the cold and hot components.
Concurrently, the upstream magnetic fields,
$\BIN$ [Figs. \ref{fig:b0}(c) and \ref{fig:b0}(d)],
are also estimated from the spectra
using Eq. (\ref{eq:b0}).
Figure \ref{fig:b0}(c) (1-mm separation) indicates the decay of the
upstream magnetic field from $t=4$ to 6 ns, 
while Fig. \ref{fig:b0}(d) (2-mm separation)
shows an initial increase from 8 to 10 ns.
In contrast, as discussed above, the spectrum at $t=12$ ns indicates
a single Maxwellian ion velocity distribution
and yields $\BIN \sim 0$, confirming that 
$\BIN$ decays completely from 10 to 12 ns.

The simultaneous estimations of the magnetic field decay and
plasma energization 
derived from the hot and cold ion components,
provide quantitative evidence of energy conversion
during the reconnection process for both inflow conditions: 1-mm and 2-mm
spot separations.
For the 2-mm separation, $\BIN$ increases during the early stage
of reconnection from 8 to 10 ns and decreases from 10 to 12 ns
as the reconnection proceeds.
Concurrently, the ion temperature increment, $\Delta\TI$, increases
from 8 to 10 ns and decreases from 10 to 12 ns.
Conversely, for the 1-mm separation, both $\BIN$ and $\Delta\TI$ decrease
from 4 to 6 ns.
This indicates that the early stage of reconnection has already passed by 4 ns
for the 1-mm separation, corresponding to the later phase observed
from 10 to 12 ns for the 2-mm separation.
Consequently, the onset of magnetic reconnection for the 2-mm separation
is delayed by more than 6 ns compared to the 1-mm separation.

This partition of dissipated magnetic energy into ion heating
and bulk kinetic energy is consistent with observations
in other laboratory plasma experiments and
kinetic simulations \cite{Ono2011-ru, Fox2011-rt}.
Particularly in collisionless or semi-collisional regimes, 
the dissipated magnetic energy is predominantly partitioned into ion thermal energy 
rather than bulk kinetic energy\cite{Yamada2014-fu, Hsu2000-dd}.
Potential explanations for this strong ion heating involve a combination of 
kinetic processes in the reconnection outflow and/or 
the specific environment of laser-driven plasmas.
For example, as cold inflow ions cross the separatrix, they may be picked up 
by the reconnected magnetic field, gaining substantial cyclotron gyromotion 
that provides an effective ion thermal energy comparable to 
the bulk kinetic energy\cite{Drake2009-hc}.
In addition, the interpenetration of counter-streaming ions and the subsequent
stagnation or thermalization via plasma instabilities and slow shocks 
can provide additional thermal energy\cite{Fox2011-rt, Hoshino1998-rt}.
However, identifying the dominant mechanism is beyond the scope of 
the present study.
Future investigations combining kinetic simulations or
higher-spatial-resolution measurements will be required to clarify
the detailed energy partition mechanisms 
in this laser-driven reconnection experiment.

\subsection{Reconnection Rate and Current Sheet Dynamics}

The magnetic reconnection rate 
can be estimated from the rate of magnetic flux
($\phi$) decay in the upstream, normalized by $\BIN$ and $\VA$:
\begin{align}
	R = \frac{1}{\VA\BIN}\left| \frac{d\phi}{dt} \right|. \label{eq:rate}
\end{align}
Here, assuming that the magnetic field in the diffusion region of 
$\delta_{\rm i}\sim\rci$
is expressed with a linear spatial distribution:
$B_x(y) = -\BIN y/\rci$ for $0\leq y \leq \rci$,
\begin{align}
	\phi \sim -\frac{1}{2}\BIN\rci.
\end{align}

\begin{figure}
\begin{center}
\includegraphics[width=\linewidth]{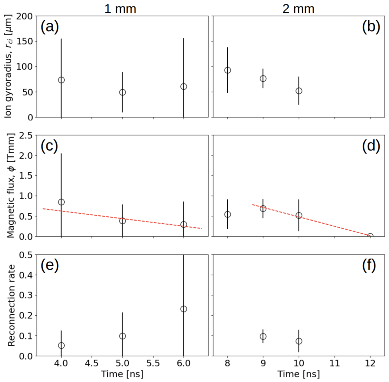}
\caption{\label{fig:rate}
Ion gyroradius, $\rci$ [(a) and (b)] for 
	the 1-mm and 2-mm separations, respectively,
	is estimated from $T_i$ and $\BIN$,
	and the upstream magnetic flux, $\phi$ [(c) and (d)],
	is estimated from $\BIN$ and $\rci$.
	The reconnection rate is calculated
	using $\BIN$ and the Alfv\'{e}n velocity $\VA$
	assuming a linear decrease of $\phi$.
}
\end{center}
\end{figure}

Following the procedure established in our previous work,
the fundamental macroscopic quantities of the reconnection region
were evaluated.
The ion gyroradius, $\rci$
[Figs. \ref{fig:rate}(a) and \ref{fig:rate}(b) for the 1-mm and 2-mm separations, respectively],
is estimated from $\TI$ and $\BIN$, and the upstream magnetic flux,
$\phi$ [Figs. \ref{fig:rate}(c) and \ref{fig:rate}(d)],
is estimated from $\BIN$ and $\rci$.
The reconnection rate is calculated from
$\BIN$ and the Alfv\'{e}n velocity $\VA$, using Eq. (\ref{eq:rate})
assuming a linear decrease of $\phi$,
and shown in Figs. \ref{fig:rate}(e) and \ref{fig:rate}(f) 
for the 1-mm and 2-mm separations, respectively.
The rate of change $d\phi/dt$ is obtained by applying a linear fit to $\phi$
from 4 to 6 ns for the 1-mm separation, and from 9 to 12 ns for the 2-mm separation
(assuming $\phi \sim 0$ at $t=12$ ns),
as shown in Figs. \ref{fig:rate}(c) and \ref{fig:rate}(d).
Note that while the increase in $\BIN$ from 8 to 10 ns 
for the 2-mm separation [Fig. \ref{fig:b0}(d)]
may be due to flux compression in the inflow, 
the magnetic flux decreases after 9 ns [Fig. \ref{fig:rate}(d)], 
suggesting magnetic dissipation in the current sheet.

Comparing the rates between the two spot separations highlights
the parameter dependence of the reconnection process,
yielding the reconnection rates that can be compared with classical
Sweet-Parker or collisionless fast reconnection models\cite{Parker1957-cs, Petschek1964, Yamada2010-jd}.
In the present analysis, a reconnection rate of 0.1--0.2 is estimated.
This value is comparable to the universal reconnection rate of 0.1\cite{Yamada2010-jd,Huba2004-tl,Cassak2017-dw,Liu2017-xe,Birn2001-mi},
and is consistent with the fact that high rates of 0.1 or even larger are
occasionally observed in laser-driven reconnection experiments\cite{Dong2012-kc, Fox2011-rt}.
Furthermore, the fast rate of $\sim$0.1 for the 2-mm separation is comparable to
that estimated in our previous research\cite{Morita2022-cm}
which was performed with the same target and 2-mm separation.

\begin{figure}
\begin{center}
\includegraphics[width=\linewidth]{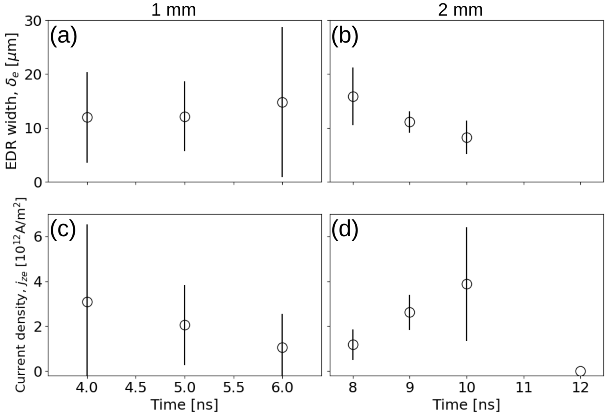}
\caption{\label{fig:jze}
	Electron diffusion region width, $\delta_{\rm e}$, estimated
	from $\TE$ and $\BIN$ for (a) the 1-mm and (b) 2-mm separations, and
	the electron current density estimated from $\BIN$ and
	$\delta_{\rm e}$
	for (c) the 1-mm and (d) 2-mm separations.
}
\end{center}
\end{figure}

Finally, the microscopic structure is characterized by estimating
the electron current density $j_{z{\rm e}}$ from $\BIN$ and
the width of the electron current sheet (electron diffusion region: EDR),
which is comparable to the scale of the electron meandering motion
expressed as\cite{Hoshino2018-wg}
$\delta_{\rm e} \sim \sqrt{\rce\rci}$,
where $\rce$ is the electron gyroradius.
The EDR widths for the 1-mm and 2-mm separations are estimated
in Figs. \ref{fig:jze}(a) and \ref{fig:jze}(b), respectively.
The EDR width for the 1-mm separation remains nearly constant from 4 to 6 ns,
while that for the 2-mm separation decreases over time from 8 to 10 ns.
This decrease suggests magnetic field compression in the inflow region;
this tendency is consistent with the ion gyroradius
[Figs. \ref{fig:rate}(a) and \ref{fig:rate}(b)]
and with the magnetic field, $\BIN$
[Figs. \ref{fig:b0}(c) and \ref{fig:b0}(d)].

By considering Ampere's law across the current sheet
(using the sheet widths shown in Figs. \ref{fig:jze}(a) and \ref{fig:jze}(b)),
the relation\cite{Morita2022-cm}
$\BIN = \mu_0 \delta_{\rm e} j_{z{\rm e}}/2$
is obtained, and $j_{z{\rm e}}$ is plotted in Figs. \ref{fig:jze}(c)
and \ref{fig:jze}(d)
for the 1-mm and 2-mm separations, respectively.
These estimated electron current densities are comparable to
those evaluated from $\text{LTS}_z$
[Figs. \ref{fig:TSz}(e) and \ref{fig:TSz}(f)],
suggesting a reasonable estimation for these reconnection parameters.
The fact that the width $\delta_{\rm e}$ is much narrower 
than the ion diffusion region $\delta_{\rm i}\sim\rci$
suggests the formation of a collisionless electron diffusion region,
which is characteristic of fast magnetic reconnection\cite{Ji2011-ew, Hesse1999-pg}.

We evaluated two different inflow conditions by varying the laser spot separation.
While the temperature is slightly higher for the 1-mm separation
compared to the 2-mm separation,
no significant differences were observed in the plasma density and $\BIN$.
In general, the Biermann battery field initially generated by laser ablation
is expected to be identical for both beam separations, and
the magnetic field at the midplane should be larger for the narrower separation.
However, the results estimated in this study
[Figs. \ref{fig:b0}(c) and \ref{fig:b0}(d)
and Figs. \ref{fig:rate}(e) and \ref{fig:rate}(f)]
reveal that magnetic diffusion into the reconnection region
does not strongly depend on the upstream magnetic field intensity
carried by each plasma bubble.
Consequently, comparable $\BIN$ and reconnection rates are obtained 
in both cases,
demonstrating a weak dependence on the initial upstream parameters.
This result suggests that the reconnection rate is less
sensitive to the global inflow conditions than to the local
physics of the current sheet and diffusion region.
The interaction geometry therefore appears to influence
the onset timing of reconnection, rather than how fast
magnetic flux reconnects once a reconnection layer is
formed.

\section{Summary}

In summary, we have presented a comparative experimental study of magnetic reconnection in laser-produced plasmas, where the reconnecting magnetic fields are self-generated via the Biermann battery effect. 
By employing collective laser Thomson scattering (LTS), we measured the scattering spectra of the reconnecting plasmas and extracted the local plasma parameters by fitting these spectra under the assumption of Maxwellian or double-Maxwellian distributions. 
This diagnostic approach allowed us to evaluate the current density
from the electron-ion drift inferred from the asymmetric LTS$_z$
spectra, while plasma heating and flow acceleration were quantified
from the hot and cold components identified in the LTS$_x$ spectra.
By combining these local measurements with macroscopic energy and mass conservation laws, we derived the upstream magnetic field, the magnetic flux, and their temporal evolution without relying on external magnetic probes. 
Through a systematic variation of the initial separation distance between the laser focal spots, we observed how the spatial scale of the colliding plasma bubbles affects the reconnection dynamics. 
Using this conservation framework, a reconnection rate of 0.1--0.2
was estimated. This value is comparable to the universal reconnection
rate of 0.1 and is consistent with the fast reconnection rates reported
in strongly driven laser experiments.

Despite the substantially different plasma expansion
histories associated with the 1-mm and 2-mm
configurations, comparable upstream magnetic fields
and reconnection rates are obtained in both cases.
This result suggests that the reconnection rate is
relatively insensitive to global inflow conditions
and is primarily determined by the local physics of
the reconnection current sheet and diffusion region.
Furthermore, our analysis of the energy partition showed that a significant portion of the dissipated magnetic energy is converted into thermal energy rather than directed bulk kinetic energy. 
The evaluation of the outflow components indicates that plasma heating exceeds bulk acceleration in this collisionless or semi-collisional regime. 

These findings provide quantitative experimental constraints on 
models of fast magnetic reconnection. 
Specifically, successful models should reproduce both the weak dependence 
of the reconnection rate on inflow conditions and the observed partition 
of dissipated magnetic energy between ion heating and bulk kinetic energy. 
Future experiments can test the validity of these constraints 
by systematically varying parameters such as guide field, 
inflow asymmetry, and Lundquist number and determining 
how the reconnection rate, current-sheet structure, 
and energy partition respond. 
Such measurements will provide direct tests of 
Hall reconnection\cite{Liu2022-gt}, 
asymmetric reconnection\cite{Cassak2007-bw}, 
guide-field reconnection\cite{Pritchett2001-mc}, 
and plasmoid-mediated reconnection\cite{Bhattacharjee2009-rq,Shibata2001-mf}, 
and help identify which models best reproduce 
both the reconnection rate and energy partition.

These constraints also provide insight into magnetic-energy dissipation 
in space and astrophysical plasmas. 
Fast magnetic reconnection is believed to play a central role 
in phenomena such as solar flares\cite{Shibata2001-mf} 
and magnetospheric reconnection\cite{Liu2022-gt}. 
Although fast reconnection is widely inferred in these systems, 
its physical origin remains an open question, 
particularly the mechanisms responsible for reconnection rates of order 0.1 
across different plasma conditions and magnetic-field configurations. 
The present constraints therefore provide a quantitative laboratory benchmark 
for testing existing theoretical and numerical models 
of fast magnetic reconnection through comparisons with 
the observed reconnection rate and magnetic-energy conversion.

\begin{acknowledgments}
	The authors would like to acknowledge the dedicated technical support
	provided by the staff at the Institute of Laser Engineering,
	the University of Osaka, for laser operation, target fabrication,
	and plasma diagnostics.
	The authors also thank Dr. N. Ozaki 
	for allowing the use of the target alignment system.
	This work was supported by JSPS KAKENHI Grant Numbers
	JP24K00605, JP24H01816, JP24K17029,
	JP23K22522, JP23H04864, JP23K20038, JP23K25907,
	JP22H00119,
	and JP20H01881;
	by the Institute of Laser Engineering
	through Joint Research Project No. 2023A1-012;
	and by the Kajima Foundation for the International Joint Research Grants (2025-06).
	Additionally, this work was the result of using research equipment
	shared in the MEXT Project for promoting public utilization
	of advanced research infrastructure
	(Program for advanced research equipment platforms)
	under Grant Number JPMXS0450300021.
\end{acknowledgments}




%

\end{document}